\documentclass[5p]{elsarticle} 
\usepackage{graphicx}
\usepackage{xcolor}
\usepackage{booktabs}
\usepackage{flushend}
\usepackage{amsmath}
\usepackage{amsthm}
\usepackage{multirow}
\usepackage{tabularx}
\usepackage{hyperref}
\usepackage{eurosym}
\newcommand{\squishlist}{
 \begin{list}{$\bullet$}
  { \setlength{\itemsep}{0pt}
     \setlength{\parsep}{1pt}
     \setlength{\topsep}{1pt}
     \setlength{\partopsep}{0pt}
     \setlength{\leftmargin}{1.5em}
     \setlength{\labelwidth}{1em}
     \setlength{\labelsep}{0.5em} } }
 \newcommand{\squishend}{\end{list}}

\begin{document}

\title{Uncovering Hidden Semantics of\\ Set Information in Knowledge Bases}


\author[add1]{Shrestha Ghosh\corref{cor1}}
\ead{ghoshs@mpi-inf.mpg.de}
\author[add1]{Simon Razniewski}
\ead{srazniew@mpi-inf.mpg.de}
\author[add1]{Gerhard Weikum}
\ead{weikum@mpi-inf.mpg.de}
 
\cortext[cor1]{Corresponding author}
\address[add1]{Max Planck Institute for Informatics, Saarland Informatics Campus, Saarbr{\"u}cken 66123, Germany}

\newcommand{\revision}[1]{{\textcolor{black}{#1}}}
\newcommand{\revisionsimon}[1]{{\textcolor{black}{#1}}}
\renewcommand{\tt}[1]{\texttt{#1}}
\renewcommand{\paragraph}[1]{\smallskip\noindent\textbf{#1.\mbox{\ \ }}}
\newtheorem{prblm}{Problem}
\newcommand{\triple}[3]{$\langle\textit{#1},\allowbreak \textit{#2},\allowbreak \textit{#3} \rangle$}
\newcolumntype{C}[1]{>{\centering}m{#1}}
\begin{abstract}
Knowledge Bases (KBs) contain a wealth of structured information about entities and predicates. This paper focuses on \emph{set-valued predicates}, i.e., the relationship between an entity and a set of entities. In KBs, this information is often represented in two formats: (i) via \emph{counting predicates} such as \tt{numberOfChildren} and \tt{staffSize}, that store aggregated integers, and (ii) via \emph{enumerating predicates} such as \tt{parentOf} and \tt{worksFor}, that store individual set memberships. Both formats are typically complementary: unlike enumerating predicates, counting predicates do not give away individuals, but are more likely informative towards the true set size, thus this coexistence could enable interesting applications in question answering and KB curation.

In this paper we aim at uncovering this hidden knowledge. We proceed in two steps. (i) We identify set-valued predicates from a given KB predicates via statistical and embedding-based features. 
(ii) We link counting predicates and enumerating predicates by a combination of co-occurrence, correlation and textual relatedness metrics. We analyze the prevalence of count information in four prominent knowledge bases, and show that our linking method achieves up to 0.55 F1 score in set predicate identification versus 0.40 F1 score of a random selection, and normalized discounted gains of up to 0.84 at position 1 and 0.75 at position 3 in relevant predicate alignments. Our predicate alignments are showcased in a demonstration system available at \url{https://counqer.mpi-inf.mpg.de/spo}. 
\end{abstract}
\maketitle              

\section{Introduction}

\begin{figure}[t]
 \centering 
 \includegraphics[width=\columnwidth]{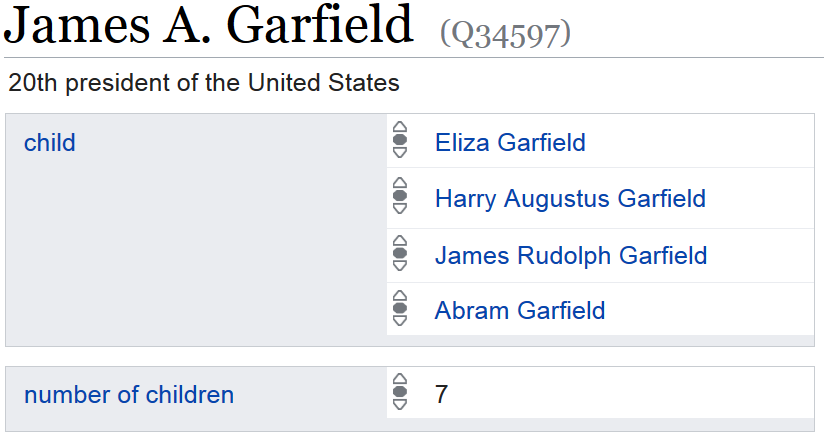}
\caption{Example of enumerating and counting predicates in Wikidata (\url{https://www.wikidata.org/wiki/Q34597}).}
 \label{fig:garfield}
\end{figure}

\paragraph{Motivation and Problem}
Knowledge bases (KBs) like Wikidata~\cite{vrandevcic2012wikidata}, DBpedia~\cite{auer2007dbpedia}, Freebase~\cite{bollacker2008freebase} and YAGO~\cite{suchanek2007yago} are important backbones for intelligent applications such as structured search, question answering and dialogue.\linebreak Properly modelling and understanding the schema of such KBs, and the semantics of their predicates, is a crucial prerequisite for utilizing them. In this paper we focus on \emph{set-valued predicates}, \emph{i.e.}, predicates which connect entities with sets of entities. Set-valued predicates typically come in two variants: (i) as enumerating predicates, which list individual objects for a given subject, and (ii) as counting predicates, which present total object counts. An example for this is shown in Fig.~\ref{fig:garfield}, an excerpt from Wikidata about the former US president Garfield. 
The predicate \tt{child}, which lists individual children of Garfield is an enumerating predicate, while \tt{numberOfChildren}, which gives a count of Garfield's children, is a counting predicate, and both model the same phenomenon. Set predicates can also model merely related phenomena, for instance, for a given location, the sets described via \tt{numberOfInhabitants} and \tt{birthPlaceOf} typically have a considerable overlap, but do not coincide. 

Identifying set predicates and set alignments would be an important step towards a better understanding of KB semantics. In particular, set alignments 
would be beneficial for the following use cases:
\squishlist
\item[(i)] {\em KB curation:} Identifying gaps and inconsistencies
in the KB and getting directives for acquiring missing pieces of knowledge
(e.g., adding the 3 absent children of US president Garfield to the KB)~\cite{mirza2018enriching}.
\item[(ii)] {\em Query formulation:} Aiding users to formulate
proper SPARQL queries by showing them related predicates
(e.g.,
finding people with more than 2 children by computing the union
of matches for the counting predicate and results from aggregating
the instances of the enumerating predicate\footnote{Example query for people with $>2$ children: \url{http://tinyurl.com/y4hemdvc}})~\cite{ontop}.
\item[(iii)] {\em Answer explanation:} Exemplifying query results
by showing key instances of queries over counting predicates
(e.g., showing a few individual Turing Award winners for a
 query about the number of award winners).
\squishend
%
\revisionsimon{Note that we do not advocate that all gaps between related predicates hint at errors or incompleteness that require actions. Scope of a KB, (non-)notability of entities, or privacy considerations may well motivate that certain gaps should not be filled, and temporal semantics may add subtleties (e.g., one predicate only counting current employees, the other storing also historical members). }

While there 
is a rich body of research
on ontology alignment and schema matching~\cite{rahm2001survey,euzenat2007ontology,shvaiko2013ontology,jain2010ontology,suchanek2011paris,wang2013unified,niepert2010probabilistic,boldyrev2018multi}, these works typically 
focus on identifying perfectly matching pairs of predicates with
the same or largely overlapping values.
This situation differs from our setting where the integer values of counting predicates and the cardinalities of enumerating predicates modelling the same or related phenomenon rarely match perfectly.
Properly identifying set predicates and set alignments in 
knowledge bases is also difficult for other reasons:
(i) KBs contain a large number of predicates, often with uninformative names and without coherent type signature, thus making the 
identification of set predicates and their alignments challenging. 
(ii) Enumerating predicates are often incomplete (like Garfield's children) and counting predicates may be approximate estimates only (like number of inhabitants); so cardinalities do not match count values, yet the predicates should be linked in order to couple them for future KB completion, consistency assessment and other use cases listed above.

\paragraph{Approach and Contribution}
This paper presents\linebreak CounQER (for ``\underline{\textbf{Coun}}ting \underline{\textbf{Q}}uantifiers and \underline{\textbf{E}}ntity-valued P\underline{\textbf{R}}edicates''), the first comprehensive methodology towards identifying and linking set predicates in KBs. CounQER is judiciously designed to identify set predicates in noisy and incomplete web-scale KBs such as Wikidata, DBpedia and Freebase. It operates in two stages: In the first stage, supervised classification combining linguistic and statistical features is used to identify enumerating and counting predicates. In the second stage, a set of statistical co-occurrence and correlation measures is used in order to link the set predicates.

Our salient original contributions are:
\begin{enumerate}
    \item We introduce the notion of set predicates, its variants, and highlight the benefits that can be derived from identifying their alignments.
    \item We present a two-stage methodology for (i) predicting the counting and enumerating predicates via supervised classification and, (ii) ranking set predicates of one variant aligned to the other variant via statistical and lexical metrics.
    \item We demonstrate the practical viability of our approach by extensive experiments on four KBs: Wikidata, Freebase, and two variants of DBpedia, .
    \item We publish results of our alignment methodology for these KBs at \url{https://tinyurl.com/y2ka4kfu} which contains 264 alignments from DBpedia mapping-based KB, 3703 alignments from the DBpedia raw KB, 25 alignments from the Wikidata-truthy KB, 274 alignments from the Freebase KB and, an interactive demo QA system at \url{https://counqer.mpi-inf.mpg.de/spo}.
\end{enumerate}

\section{Related Work}

\paragraph{Schema and ontology alignment}
Schema alignment is a classic problem in data integration~\cite{rahm2001survey}. For ontologies and on the semantic web, added complexity comes from taxonomies and ontological constraints~\cite{euzenat2007ontology,shvaiko2013ontology}. Approaches to ontology alignment include BLOOMS~\cite{jain2010ontology} and PARIS~\cite{suchanek2011paris}, voting-based aggregation~\cite{wang2013unified}, probabilistic frameworks~\cite{niepert2010probabilistic}, or methods for the alignment of multicultural data~\cite{boldyrev2018multi}. These methods typically rely on a combination of lexical, structural, constraint and instance based information. Some works have also investigated subset relations~\cite{koutraki2017online}, yet still focusing only on entity-entity relations.
The most important venue in the field of ontology alignment is the long-running Ontology Matching workshop series~\cite{ontology-workshop} along with its attached challenges~\cite{ontology-challenge}.
Our setting, where enumerations need to be aligned with counts, is atypical in ontology alignment and has not received prior attention.

\paragraph{Set information in logics and KBs}
Modelling count information has a history in qualifying number restrictions~\cite{hollunder1991qualifying} and role restrictions in description logics~\cite{calvanese1998description}. In the OWL standard~\cite{mcguinness2004owl}, count information on relations can be expressed via cardinality assertions. The second statement in Fig.~\ref{fig:garfield}, for instance, could be expressed as

\medskip

\texttt{\hspace{-0.3cm}ClassAssertion(ObjectExactCardinality(7 :child) :Garfield)}

\medskip\noindent
OWL furthermore also supports lower bounds, e.g., that a certain person has at least two children, and upper bounds, e.g., that a certain car has at most five seats. The added expressiveness from counting quantifiers typically comes at a complexity tradeoff, no matter whether they are part of the ontology language~\cite{glimm2008conjunctive,calvanese2009regular}, or only of the query language~\cite{fan2016adding,nikolaou2019foundations,dell2019counting}, especially as they introduce negation (\emph{``Companies that are owned by zero other companies''}).
\revision{Despite several methods for automatically learning logical axioms and patterns~\cite{lehmann2010,galarraga2013amie}, we are not aware of attempts to learn set relatedness.}

\paragraph{Numeric and set information in KBs, QA and IE}
Popular knowledge bases contain considerable numeric information. Research has focused on detecting errors and outliers in such information~\cite{wienand2014detecting}, and in organizing and annotating measurement units~\cite{neumaier2016multi,subercaze2017chaudron}.

Set information is important for question answering. For instance, in~\cite{mirza2018enriching} it is reported that between 5\% and 10\% of questions in popular TREC QA datasets concern counts. This information need\revision{, although acknowledged by QA systems, is so far not dealt with in a principled manner}. AQQU~\cite{bast2015more}, for instance, includes a special translation for questions starting with \emph{``How many?''}. The Google search engine similarly answers count keyword queries for popular entities like \emph{``How many children Angelina Jolie''} with both counts and instances, \revision{yet there is no discernible pattern concerning which queries evoke this behaviour, and it occurs only for very few highly popular ones}.

When concerned with numeric information, textual information extraction traditionally focused on temporal information~\cite{ling2010temporal} and measures~\cite{saha2017bootstrapping}. Recently also counting information extraction has received attention, e.g., from sentences like \textit{``The LoTR series consists of three books''}~\cite{mirza2017cardinal,mirza2018enriching}. Such information can then be used to assess and improve KB completeness~\cite{razniewski2016but,razniewski2019coverage}. \revision{ In the present work, we investigate the identification and alignment of set predicates in knowledge bases itself, i.e., without external text sources.}

\paragraph{KB recall information} Recall is an important dimension of KB quality, with impact on downstream use cases~\cite{paulheimknowledgebase}. Unlike precision, it cannot be easily evaluated by sampling. Existing approaches to KB recall estimation can be grouped into three families. The first are approaches based on statistical patterns in the data, e.g., sample overlap~\cite{speciessamplingcompleteness}, digit-distributions~\cite{Soulet2018RepresentativenessLaw}, or association rules~\cite{Galarraga2017PredictingBases}. The second are relative approaches, i.e., where recall expectations are collected from related entities~\cite{Hopkinson2018Demand-WeightedBase, Soulet2018RepresentativenessLaw}. The third are text-extraction based approaches mentioned above. \revision{ The present work complements these approaches with the exploration of predicate interrelations among set predicates.}

\section{Problem Definition}
\label{sec:problem}

Let $P$ be a set of predicates. A \textit{knowledge base} (KB) is a set of triples $(s,p,o)$, where $p\in P$, $s$ is an entity, and $o$ is either an entity or a literal. For the remainder of this paper we assume that each triple $(s,p,o)$ with an entity as object also exists in its inverse form $(o,p^{-1},s)$ in each KB, thus the following elaborations need to consider only one direction.


The foundational concept for this work is that of a set predicate.
\theoremstyle{definition}
\newtheorem{definition}{Definition}
\begin{definition}[Set Predicate\footnote{\revisionsimon{We emphasize that \emph{set predicate} refers to the intended semantics of the modeller, not to be mixed with the capabilities of the modelling language. In particular, unlike SQL, the RDF data model does not know a \emph{SET} datatype, but can capture sets via multiple triples sharing subject or object.}}]
A predicate which \emph{conceptually} models the relation between an entity and a set of entities is a set predicate.
\end{definition}

Set predicates can be expressed in KBs in two variants: Via binary predicates that enumerate individual set members, and via counting predicates that abstract away from individuals, and store aggregate counts only.

\begin{definition}[Enumerating Predicate]
A set predicate that models sets via binary membership statements is called an enumerating predicate.
\end{definition}

\newcommand{\KB}{\mathit{KB}}

We denote the set of enumerating predicates as $E$. 


\begin{definition}[Counting Predicate]
\revisionsimon{A set predicate whose values represent counts of entities modelled in the considered KB is called a counting predicate.}
\end{definition}

Entity counts necessarily are integers, yet KB predicates can contain integers that represent a variety of other concepts, for instance identifiers or measures like length and weight.


Following these definitions, the predicates \tt{child} and\linebreak \tt{numberOfChildren} in Fig.~\ref{fig:garfield} are set predicates, the former an enumerating predicate, the latter a counting predicate. Other examples are \tt{worksAt$^{-1}$} and \tt{authorOf}, which frequently enough take several values for a subject. This is in contrast to predicates of a functional or quasi-functional nature, such as \tt{bornIn} and \tt{mother}, which predominantly take a single object, and hence, where counts are uncommon and rarely informative. Yet the threshold for enumerating predicates is imprecise, for instance, the predicate \tt{citizenOf} is predominently functional, but some entities still have multiple citizenships which are conceivably countable.

Other examples of counting predicates are \tt{popula\-tion}, \tt{numberOfStudents} and \tt{airlineDestinations}. The distinction between counting predicates and measurement predicates like \tt{riverLength} and \tt{revenue} is quite crisp, since measurements usually come with units (km, \euro, etc.) and can take fractional values ($1.7$ km) while entity counts cannot. Our definition is phrased to also exclude some predicates taking integer values, like \revision{\tt{episode\-Number} (not a count, but a sequential number assigned to an episode of a TV series)} and \tt{floor\-Count} (a count, but not of something commonly considered as entities). Thus, integer values are a necessary but not a sufficient condition for being a count predicate. 

We summarize our first problem as follows.

\begin{prblm}[Set predicate identification]
Given a \textit{KB} with a predicate set $P$, identify the set of enumerating predicates, $E$, and the set of counting predicates, $C$.
\end{prblm}

Note that the above definitions are conceptual only. Functionalities computed over actual KBs are unreliable due to incompleteness, errors, and redundancies, and common KBs do not have a \textit{entity-count} datatype. Thus, in later sections we will develop supervised classifiers for identifying both kinds of set predicates.

Let us now turn to the relation between set predicates. 
\revisionsimon{\textit{Set-relatedness} refers to the (ideal) amount of overlap that two set predicates have on a per-subject basis. For instance, in a perfect KB, \tt{child} and \tt{numberOfChildren} describe exactly the same set of objects per subject, once via a listing of names, ones via the aggregate count.}

\revisionsimon{In turn, the predicates \tt{population} and $\tt{bornIn}^{-1}$ do not describe the same sets, but would typically exhibit a significant overlap (many people live in the same place they are born in, though neither entails the other).
On the other hand, \tt{population} and $\tt{headquarterLocation}^{-1}$ are not set related. Although population sizes and the number of company headquarters in a place are correlated numbers, the described entities do not overlap at all, instead, are even of distinct types (\tt{person} and \tt{company}).}

\revisionsimon{Conceptually, set relatedness between two predicates can therefore be computed as an across-subjects aggregated set overlap measure, with perfect matches being the strongest relatedness. Note that this definition is strictly conceptual, since in actual KBs, counting predicates do not give away which actual entities they count.}



\begin{prblm}[Set predicate alignment]
Given sets of enumerating predicates $E$ and counting predicates $C$, for each set predicate $p\in E \cup C$, rank the predicates from the other set by their set-relatedness.
\end{prblm}

Note that the above definitions of set-relatedness are conceptual definitions. In practice, KBs do not give access to the entities counted by counting predicates, instead one only sees aggregate counts. To quantify and qualify set-relatedness, in the following sections, we will thus build a set of unsupervised alignment heuristics.

\section{Design Space and Architecture}

\paragraph{Design space}
Our goal is to develop a robust set predicate identification and linking methodology, that, with limited supervision, can work across different KBs.

If knowledge bases were clean, set predicate identification could solely rely on relation functionality and data\-types. As this is not the case in practice~\cite{wu2014redundant,wienand2014detecting,zaveri2016quality}, we instead propose to approach set predicate identification via a supervised classification framework that combines a diverse set of textual, datatype, and statistical features.
Schema alignment can also in principle be approached via hand-crafted rules, heuristic alignment metrics, or supervised learning. Due to the particularities of individual predicates and KBs (most set predicates have only very few good alignments), to avoid overfitting, we opt here for a set of heuristic alignment metrics, which we design in order to capture various desiderata of meaningful alignments, and combine them in an ensemble metric.

\begin{figure}[t]
 \centering 
 \includegraphics[width=\columnwidth]{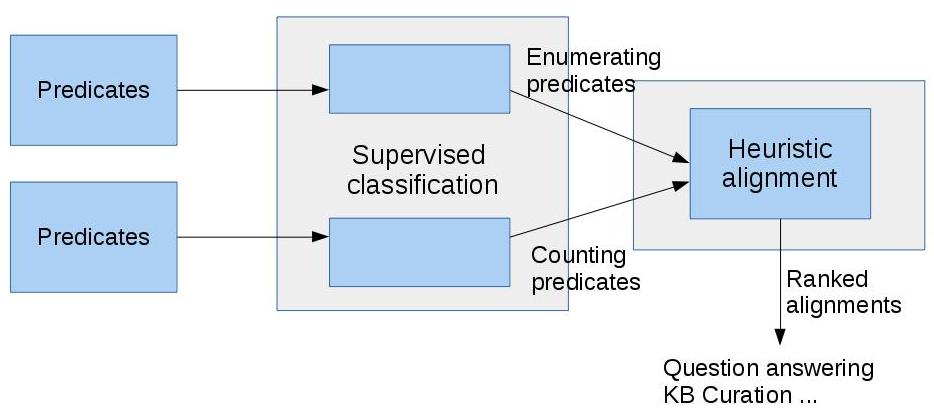}
\caption{Architecture of the CounQER approach.}
 \label{fig:architecture}
\end{figure}

\paragraph{Architecture}
Following the above considerations, we split our CounQER methodology in two steps: (i) supervised predicate classification, and, (ii) heuristic predicate alignment (see Fig.~\ref{fig:architecture}).

In the first phase, \emph{supervised predicate classification}, we use two classifiers to predict the two set variants, namely, enumerating and counting predicates. We rely on a family of features, most importantly, (i) set-related textual features, extracted from a background corpus, (ii) type information about the domain and range of the predicates, and (iii) statistical features about the number of objects per subject at different percentiles. 

In the second phase, \emph{heuristic predicate alignment}, we identify related counting and enumerating predicates using (i) set predicate co-occurrence information, (ii) set predicate value distribution, and (iii) linguistic relatedness. By assigning each pair of enumerating and counting a relatedness score, we can rank related predicates accordingly. While we evaluate the heuristics on labelled data, they are highly complementary, and thus, the choice of the heuristic to be used can be adapted to particular use cases.


\paragraph{KB assumptions}
Our approach is designed to work on a variety of knowledge bases, without requiring strong assumptions on their internal structure. Fulfilment of the following features is desirable, though not essential: (i) High-level categories/classes for entities, in particular \emph{Person, Place, Organisation, Event} and \emph{Work}. Where these are not available, we utilize links to Wikidata to extract them. (ii) High-level datatypes, in particular \emph{float}, \emph{int} and \emph{date}. Where these are not available, we utilize standard parsers and simple heuristics, such as that numbers between 1900 and 2020 are likely dates. (iii) human-readable labels for properties, with spaces or in camel case notation. Where these are not available, we deactivate corresponding linguistic features.

\section{Set Predicate Identification}
\subsection{Enumerating Predicates}

As stated in Section.~\ref{sec:problem}, if KBs were clean, functionality (\#triples per subject) would be the criterion for identifying enumerating predicates.

Yet actual KBs contain a considerable amount of noise, are incomplete, and blur functionality by redundancies (e.g., listing both the birth city and country of a person under \tt{birthPlace}). In CounQER, we thus rely on supervised classification, where functionality is only one among several features towards enumerating predicate identification.



\subsection*{a. Textual features}
Where KBs use human-readable predicate names, a basic sanity check for enumerating predicates is to verify whether in human language, the predicate name is used both in singular and plural.

\begin{enumerate}
    \item \textit{Plural-singular ratio:} For each predicate, we apply a heuristic to we generate its plural/singular form. 
    First we identify the last noun in the predicate label using the Python \textit{nltk} package and then we use the Python \tt{inflect} library to identify its form (singular/plural) and convert it to the other (plural/singular). We then compute the text frequency ratio based on the Bing API, obtaining, for instance, for $\frac{|\tt{children}|}{|\tt{child}|}$ a ratio of $\frac{128.000.000}{87.000.000}=1.47$, while for \tt{birthplace}, the ratio is $\frac{|\tt{birthplaces}|}{|\tt{birthplace}|}=\frac{1.550.000}{21.000.000}$, a ratio of 0.08.
 \end{enumerate}  

\subsection*{b. Type information}
\revision{Subject and object types may explain the applicability of other features, and certain types of objects may more naturally be counted than others, and certain subjects may more frequently come with set predicates than others.  To avoid overfitting and ensure compatibility across KBs, in this work we only consider five frequent and general classes, \{\emph{Person}, \emph{Place}, \emph{Organisation}, \emph{Event} and \emph{Work}\}, which we use to capture the domain and range type of a predicate. Further details follow in Sec.~\ref{sec:preproccessing}.} 
\begin{enumerate}
    \setcounter{enumi}{1}
    \item \textit{Predicate domain:} We encode the most frequent class of a predicate domain via binary features per class, including a 6th class for \emph{other}.
    \item \textit{Predicate range:} We encode the class of predicate range in binary variables (same as in predicate domain). 
\end{enumerate}

\subsection*{c. KB statistics}
KB statistics instantiate the observed functionality. As functionality may be blurred by outliers or a long tail of single-valued subjects, we input various datapoints in order to increase resilience of the measure. We also include basic information on datatypes.
\begin{enumerate}
    \setcounter{enumi}{3}
    \item \textit{Mean, maximum, minimum, 10th and 90th percentile number of objects per subject (functionality):} These features describe the number of objects a predicate takes per subject, with mean and percentiles giving resilience against rare outliers. For example, \tt{occupation} in Wikidata-truthy KB has a mean of $1.3$, maximum of 30, minimum of 1, 10th percentile of 1 and 90th percentile of 2. The predicate \tt{placeOf\linebreak~Birth} in Wikidata-truthy KB has a maximum of 6 objects per subjects and 1 object per subject for the other features - minimum, mean, 10th and 90th percentile.  
    \item \textit{Datatype distribution:} The fraction of triples of a predicate whose objects are of datatype entity. 
    For instance, both the predicates \tt{occupation} and \tt{place\-OfBirth} take entities for $99\%$ of the triples in Wikidata-truthy KB.  

\end{enumerate}

These features are then used for a binary classifier.

\subsection{Counting Predicates}
As per our conceptual definition, counting predicates are distinguished by having \textit{entity-count} as their datatype. As none of the KBs investigated in this paper records such a datatype, we have to use various heuristic towards identifying counting predicates. An important necessary condition are integer values, yet these alone are not sufficient. We utilize the following classes of features.

\subsection*{a. Textual features}

\begin{enumerate}
    \item \textit{Plural-singular ratio:} This feature captures the plural/singular ratio of a predicate obtained exactly as for enumerating predicates. 
\end{enumerate}

\subsection*{b. Type information}

\begin{enumerate}
    \setcounter{enumi}{1}
    \item \textit{Predicate domain:} We identify the domain of the predicates by tracing the class of the predicates to one of the most general classes in the type heirarchy, \{\textit{Place}, \textit{Person}, \textit{Organization}, \textit{Event}, \textit{Work}\}. Each of the domain class is encoded as a binary variable in the classifier. 
\end{enumerate}

\subsection*{c. KB statistics}
\begin{enumerate}
    \setcounter{enumi}{2}
    \item \textit{Datatype distribution:} We calculate the fraction of triples of a predicate taking integer values over the total number of triples of that predicate. For instance, the predicate \tt{numberOfEpisodes} in the DBpedia mapping-based KB takes only integer values, whereas \tt{episodeNumber} in the DBpedia raw KB takes integer values for $96\%$ of the triples.     
    \item \textit{Mean, maximum, minimum, 10th and 90th percentile of count value:} These features describe the actual integer value of the predicate, e.g., the mean for \tt{numberOfEpisodes} (DBpedia mapping-based KB) is 106, the maximum is 90015, the minimum is 0, the 10th percentile is 6 and the 90th percentile is 156. 
    \item \textit{Mean, maximum, minimum, 10th and 90th percentile of the number of objects per subject (functionality):} These features describe the number of integer valued triples per subject.
    
    For example, the mean \tt{numberOfEpisodes} (DBpedia-mapping-based KB) a subject takes is 1, the maximum is 8, the minimum, the 10th percentile and the 90th percentile all are 1, \emph{i.e.,} most subjects have only one fact containing this predicate. 
    In contrast, an ordinal integer predicate like \tt{episodeNumber} (DBpedia raw KB) has the following statistics - mean 32, maximum 975, minimum 1, 10th percentile 6 and 90th percentile 66. This odd behaviour is exhibited because the article page lists all or a subset of the episode numbers in a series~\footnote{DBpedia subjects with count of \tt{episodenumber} facts \url{https://tinyurl.com/dbpedia-raw-episodenumber}}.
\end{enumerate}

\section{Heuristic Predicate Alignment}
The output of the previous stage are the enumerating predicates $E$ and the counting predicates $C$. The task of this stage is to find for each predicate in $E\cup C$ the most set-related predicates from the other set. As this task may to some extent be KB-specific, we approach it via a set of unsupervised ranking metrics. We introduce three families of metrics for predicate pairs: (a) set predicate co-occurrence, based on the number of subjects for which $e \in E$ and $c \in C$ co-occur, (b) set predicate value distribution, based on the relation between the number of objects in $e$ and the value of $c$ for co-occurring subjects, and (c) set predicate linguistic similarity which measures the relatedness between the labels of the set predicates $e$ and $c$. \revision{Here, we illustrate each heuristic using the enumerating predicate \texttt{gold$^{-1}$}, which links an entity (sportsperson) to the winning sports events, and the counting predicate \texttt{singlestitles}, which gives the count of singles titles of a tennis player.}

\paragraph{A. Set Predicate Co-occurrence} \label{par:metric_cooccur}
Our first family of heuristics ranks predicates by their co-occurrence. Co-occurrence is an indication towards topical relatedness, and we propose various measures that capture absolute and relative co-occurrence frequencies. 

\begin{enumerate}
    \item $\textit{Absolute}(e,c)$: The number of subjects which have triples with both $e$ and $c$ set predicates. For instance, 
    \begin{multline*}
    \revision{\textit{Absolute}(\tt{singlestitles,gold$^{-1}$}) = 64.}
    \end{multline*}
    
    \item $\textit{Jaccard}(e,c)$: The ratio of the absolute number of subjects for which $e$ and $c$ co-occur, \emph{i.e.,} $\textit{Absolute}(e,c)$ divided by the union of subjects which take either $e$ or $c$ or both. For instance,
    \begin{multline*}
    \revision{\textit{Jaccard}(\tt{singlestitles,gold$^{-1}$})=0.006.}   
    \end{multline*} 
    \item $\textit{Conditional}(e,c)$: Co-occurrence can also be expressed as a conditional probability, i.e., the ratio of the absolute value, $\textit{Absolute}(e,c)$, to the number of subjects which take \revision{either} $c$ or $e$. For our given example,
    \revision{
    \begin{multline*}
        \textit{Conditional}_\textit{E}(\tt{singlestitles,gold$^{-1}$}) \\= 0.011.
    \end{multline*}}
    \revision{with respect to subjects only taking the predicate \tt{gold$^{-1}$} and,}
    \begin{multline*}
        \revision{\textit{Conditional}_\textit{C}(\tt{singlestitles,gold$^{-1}$}) = 0.015.}
    \end{multline*}
    \revision{with respect to subjects only taking the predicate \tt{singlestitles}.
    This implies that if a given subject has the predicate \tt{singlestitles}, it is more likely that the subject also has the predicate \tt{gold$^{-1}$} than the other way around.}
    \item $\textit{PairwiseMutualInformation}(e,c)$ or (\textit{P'wiseMI}): The log of the ratio of the joint distribution of $e$ and $c$ to their individual distributions.
    \revision{\begin{multline*}
        \textit{PMI}(e,c) = \\\log_2 \frac{|\{s \mid s \in \text{\triple{s}{e}{$\cdot$}}; s \in \text{\triple{s}{c}{$\cdot$}} \}| \times |\{s \mid s \in \text{\triple{s}{$\cdot$}{$\cdot$}}\}|}{|\{s \mid s \in \text{\triple{s}{e}{$\cdot$}}\}| \times |\{s \mid s \in \text{\triple{s}{c}{$\cdot$}}\}|}
    \end{multline*}}
    \begin{multline*}
        \revision{\textit{PMI}(\tt{singlestitles,gold$^{-1}$}) = -5.2.}
    \end{multline*}
    \revision{which implies that the two predicates are less likely to co-occur than expected from their individual occurrences.} In general, this metric ranges between $-\infty$ and $\min(-\log p(e), -\log p(c))$. The lower bound is reached when the pair $(e,c)$ does  not co-occur for any subject and the upper bound is reached when either $e$ always co-occurs with $c$ or vice-versa.

\end{enumerate}

\paragraph{B. Set Predicate Value Distribution}\label{par:metric_vald}
Co-occurrence is important but can nonetheless be spurious, e.g., when many sports teams have both the predicates \tt{stadiumSize} and \tt{coachOf}. A possibly even stronger indicator for set relatedness is a match or correlation in values, i.e., if across subjects, the number of values for the enumerating predicate, and the count stored in the counting predicate, coincide, or correlate. We propose three variants: (\ref{item:pmr}) To count the number of exact matches, and in (\ref{item:corr}) and (\ref{item:ptile}) two relaxed metrics that look for correlation and percentile similarity.

\begin{enumerate}
    \setcounter{enumi}{4}
    \item $\textit{PerfectMatchRatio}(e,c)$: The ratio of subjects where the number of objects in $e$ exactly matches the value in $c$ to the number of subjects which takes both $e$ and $c$ predicates. \revision{For example,}
    \begin{multline*}
        \revision{\textit{P'fectMR}(\tt{singlestitles,gold$^{-1}$}) = 0.125.}
    \end{multline*} \label{item:pmr}
    \item $\textit{Correlation}(e,c)$: The Pearson correlation between the size of objects of $e$ and the value of $c$ for all subjects in which they co-occur. \revision{For the above predicate pair,}
    \revision{\begin{multline*}
        \textit{Correlation}(\tt{singlestitles,gold$^{-1}$}) = 0.724.
    \end{multline*}}\label{item:corr}
    \item $\textit{PercentileValueMatch}(e,c)$: A softer score than a perfect match ratio (B~\ref{item:pmr}), for matching the $90$th percentile value of the number of objects that $e$ takes per subject with the $90$th percentile value of the $c$, such that the closer the value is to 1 the better the alignment. Let $O_c$ and $O_e$ denote the distribution of the values and the \#objects per subject, respectively.
\begin{multline*}
    \textit{P'tileVM}(e,c) = \\\textit{Min}\bigg(
    \frac{\textit{p'tile}(O_{e}, 90)}{\textit{p'tile}(O_{c}, 90)},
    \frac{\textit{p'tile}(O_{c}, 90)}{\textit{p'tile}(O_{e}, 90)}\bigg)
\end{multline*}
    \begin{multline*}
       \revision{\textit{P'tileVM}(\tt{singlestitles,gold$^{-1}$}) = 0.333.}
    \end{multline*}\label{item:ptile}

\end{enumerate}

\paragraph{C. Linguistic Similarity}\label{par:metric_ling}
Besides co-occurence, also correlations can be spurious. For instance, \tt{population} and \tt{headquarterLocation}$^{-1}$ are well correlated (bigger cities host more companies), but nonetheless, they refer to completely different kinds of entities (persons vs.\ companies). Our third family of heuristics thus looks at topical relatedness.
\begin{enumerate}
    \setcounter{enumi}{7}
    \item $\textit{CosineSimilarity}(e,c)$ measures the cosine of the angles between the average of the sets of word vectors of the labels of $e$ and $c$ obtained from pre-trained Glove embeddings~\cite{penningtonSM14} using the Python \textit{Gensim} library. 
    Wikidata predicate labels are already individual words, for DBpedia and Freebase we split the predicates at capitalization and punctuation, respectively. \revision{For example, \tt{headquarterLocation$^{-1}$} becomes \{\tt{headquarter, location}\} and \tt{race\_count} becomes \{\tt{race}, \tt{count}\}. }
\begin{multline*}
    \revision{\textit{CosineSim}(\tt{titles,gold$^{-1}$}) = 0.318}
\end{multline*}
    \revision{Out of vocabulary words, like \tt{singles\-titles}, lead to an empty word list. Similarity with an empty word list is assigned a score of zero as follows.} 
\begin{multline*}
    \revision{\textit{CosineSim}(\tt{singlestitles,gold$^{-1}$}) = 0.} 
\end{multline*}
\end{enumerate}

\paragraph{D. Alignment Summary}
\revision{In the following experiments we evaluate the alignment heuristics individually against ground truth annotations on the NDCG~\cite{jarvelin2002cumulated} score.
In this way, we can discover the best performing heuristic, and with enough training data, could even perform ensemble learning. Yet as reliance on a single heuristic would be brittle, and ensemble learning requires larger evaluation data, as a robust best-effort, we propose here to retain from each of the three families of heuristics the best performing one, and merge their scores by using averaging.}



\section{Experiments}

\subsection{KBs used}
We use four popular general purpose KBs: (i) DBpedia raw extraction~\cite{auer2007dbpedia}, (ii) DBpedia mapping-based extraction\footnote{\revision{We used DBpedia version 2016-10 for both extraction.}}~\cite{lehmann2015dbpedia}, (iii) Wikidata truthy\footnote{\revision{We used the version as of the Oct 2018.}}~\cite{vrandevcic2012wikidata} and (iv) Freebase\footnote{\revision{We used the version as of 2019 July.}}~\cite{bollacker2008freebase}. We analyze each KB in terms of predicate coverage.
\begin{enumerate}
    \item \textbf{DBpedia raw} \emph{(52.6M triples)}\textbf{:} All predicate-value pairs present in the infoboxes of English Wikipedia article pages.
    \item \textbf{DBpedia mapping-based} \emph{(29M triples)}\textbf{:} A cleaner infobox dataset where predicates were manually\linebreak mapped to a human-generated ontology. Unmapped predicates and type violating triples are discarded.
    \item \textbf{Wikidata truthy} \emph{(210.3M triples)}\textbf{:} 
    Simple triple export of Wikidata that ignores some advanced features such as qualifiers and deprecated ranks. 
    \item \textbf{Freebase} \emph{(1B triples)}\textbf{:} The tuple store available as an RDF dump at \url{https://developers.google.com/freebase}.
\end{enumerate}

We also analysed YAGO~\cite{suchanek2007yago} \emph{(1.1B triples)}, a WordNet-aligned and sanitized harvest of Wikipedia infobox statements, containing only 76 distinct predicates. By manual inspection we found several enumerating predicates, like \tt{hasChild} and \tt{is\-CitizenOf}, but only one counting predicate, \tt{numberOfPeople} and therefore refrained from further processing of this KB.
\begin{table}[t]
    \centering
    \begin{tabular}{|c|c|c|}
        \hline
        \textbf{KB} & \textbf{All} & \textbf{Frequent} \\\hline
        DBP-raw & 73,234 & 16,635 \\
        DBP-map & 2,008 & 1,670 \\
        WD-truthy & 6,111 & 4,067 \\ 
        Freebase & 799,807 & 13,872\\ 
        \textit{YAGO} & \textit{(79)} & \textit{(79)} \\ \hline
    \end{tabular}
    \caption{Total number of KB predicates (direct + inverse) and most frequent ones.}
    \label{tbl:freq_predicates}
\end{table}

On adding inverse triples, i.e., adding $(o,p^{-1},s)$ for every $(s,p,o)$ where $o$ is an entity, the size of DBpedia-raw increased by 7.6M, DBpedia-map by 18M, Wikidata by 101.1M and Freebase by 442.1M. 

To reduce noisy data we use predicates which appear in at least 50 triples. In Table~\ref{tbl:freq_predicates} we show the number of predicates that remain after filtering all infrequent predicates. It is evident that the cleaner KBs like Wikidata and DBpedia mapping-based KB have better predicate representation. Freebase and DBpedia raw KBs are noisier with a very long tail of less frequently occurring predicates.

\subsection{Preprocessing}\label{sec:preproccessing}

\paragraph{Predicate statistics computation}
Given a KB of SPO triples, we generate the descriptive statistics of the KB predicates including (i) the datatype distribution - fraction of the triples of a predicate which take \textit{integer}, \textit{float}, \textit{date}, \textit{entity} and \revision{comma-separated} \textit{string} values, (ii) the mean, maximum, minimum, 10th and 90th percentile of the integer values that a predicate takes, (iii) the mean, maximum, minimum, 10th and 90th percentile of the number of entities per subject of a predicate and, (iv) the mean, maximum, minimum, 10th and 90th percentile of the number of integer values per subject of a predicate. \revision{We identify comma-separated string values as a datatype in order to handle noisy representations, especially in DBPedia-raw, where object sets are often captured in a single string with comma separation (e.g., ``children: Mary, John, Susan'').} 

\paragraph{Type information}
We then proceed to find the predicate domain and range. To maintain uniformity across KBs we trace the type to one of the more general classes in the type hierarchy, \{\textit{Place, Person, Organization, Event, Work}\}, \revision{with the default fallback class for entities being \textit{Thing} and non-entities being \textit{Literal}. The fallback classes capture long-tailed classes and string objects, which have no class information.} 

We sampled $100$ subjects and $100$ objects for each predicate and selected the majority class in each set as the domain and range of the predicate. \revision{Across all four KBs, the (micro-average) coverage of the predicate domain by the classes are \{\textit{Place: $18\%$, Person: $23\%$, Organization: $14\%$, Event: $4.75\%$, Work: $17.25\%$, Thing: $23\%$}\} and, for predicate range, \{\textit{Place: $18.25\%$, Person: $20\%$, Organization: $15.25\%$, Event: $1.25\%$, Work: $17\%$, Thing: $1.75\%$, Literal: $26.5\%$}\}}.

\paragraph{Linguistic features}
The frequency of occurrence of a predicate in the web in singular and plural form is determined from the total estimated web search matches returned by the Bing custom search API\footnote{https://azure.microsoft.com/en-us/services/cognitive-services/bing-custom-search/}. \revision{For inverse predicates, we reuse the predicate labels of their forward form for getting the textual features for the classifiers and the linguistic similarity measure for the alignment heuristics.}

\subsection{Training and evaluation data}


We prepared the data for the \textit{classification step} by employing crowd workers to annotate $400$ randomly selected \revision{predicates} for enumerating predicates and $400$ for counting predicates from the four KBs - taking 100 from each KB. The annotation task comprised a predicate and five sample subject-object pairs with options to select if the predicate was likely a set predicate (enumerating or counting).

An example question for annotating counting predicates is given below.

\smallskip

\textit{\textbf{Q:} Based on the following facts, decide whether the relation gives a count of unique entities.}

    \begin{tabular}{lll}
        The Herald (Sharon) & circulation & 15715 \\
        H.O.W. Journal & circulation & 4000\\ 
        L'Officiel & circulation & 101719\\
        The Music Scene (magazine) & circulation & 25000\\
        Pipe Dream (newspaper) & circulation & 7000\\
    \end{tabular}
    
\smallskip
    
\noindent\textit{Options:} $\circ\textit{Yes}$ $\circ\textit{Maybe yes}$ $\circ\textit{Maybe no}$ \revision{$\circ\textit{No}$} $\circ\textit{Do not know}$

\medskip
\noindent
For \revision{enumerating} predicate annotation we used the following question format.

\smallskip

\textit{\textbf{Q:} Based on the following facts, decide whether the relation enumerates entities.}
\smallskip

\scalebox{0.85}{\hspace{-0.7cm}
\begin{tabular}{lll}
     A Low Down Dirty Shame & producer & Mike Chapman \\
     Bye Bye Brazil & producer & Luiz Carlos Barreto\\
     Heaven Knows, Mr. Allison & producer & Eugene Frenke\\
     Surviving Paradise & producer	& Kamshad Kooshan\\
     I'll Come Running Back to You & producer & Bumps Blackwell\\
\end{tabular}
}

\smallskip
\noindent\textit{Options:} $\circ\textit{Yes}$ $\circ\textit{Maybe yes}$  $\circ\textit{Maybe no}$ \revision{$\circ\textit{No}$} $\circ\textit{Do not know}$

\medskip
\noindent

We collected three judgements per predicate, \emph{i.e.,} a total of $2400$ annotations (2 variants of set predicates $\times$ 4 KBs $\times$ 100 predicates $\times$ 3 judgments). The options in the annotation task are graded. We translated the labels into numeric scores \{\textit{Yes}: 1, \textit{Maybe yes}: 0.75, \textit{Do not know}: 0.5, \textit{Maybe no}: 0.25, \textit{No}: 0\}, with the final label being the average of all judgments. \revision{Concerning annotator agreement, we found the pooled standard deviation of the scores per predicate to be $0.41$. We only keep rows with a clear polarity, \emph{i.e.,} rows with average score outside the interval $(0.4, 0.6)$, effectively excluding rows averaging around the option \textit{Do not know}. The labels of the remaining rows are then translated into binary 0-1-judgments. Of the counting predicate rows, $86.25\%$ showed a clear polarity, of the enumerating predicate rows, $82\%$.}
Thus, we obtained our training data, with 39 positive and 306 negative data points for the counting classifier and, 133 positive and 195 negative data points for the enumerating classifier.



\begin{table}[t]
    \centering
    \begin{tabular}{|l c c|}
    \hline
         \textbf{KB} & \textbf{Positive samples} & \textbf{Negative Samples}  \\
         \hline
         DBP-raw & 16 & 62\\
         DBP-map & 9 & 72\\
         WD-truthy & 7 & 87\\
         Freebase & 7 & 85\\
         Total & 39 & 306 \\
         \hline
    \end{tabular}
    \caption{Distribution of counting classifier training samples across KBs.}
    \label{tbl:perkb-cp}
\end{table}

\begin{table}[t]
    \centering
    \begin{tabular}{|l c c|}
    \hline
         \textbf{KB} & \textbf{Positive samples} & \textbf{Negative Samples}  \\
         \hline
         DBP-raw & 33 & 53\\
         DBP-map & 27 & 58\\
         WD-truthy & 27 & 55\\
         Freebase & 46 & 29\\
         Total & 133 & 195 \\
         \hline
    \end{tabular}
    \caption{Distribution of enumerating classifier training samples across KBs.}
    \label{tbl:perkb-ep}
\end{table}

\revision{We can conclude from Table~\ref{tbl:perkb-cp} that in general, KBs contain comparably few counting predicates, which also contributes to the low precision score of the counting classifier. From the numbers in Table~\ref{tbl:perkb-ep}, we observe that enumerating predicates have a comparably higher occurrence.}

For the \textit{alignment step}, evaluation data was prepared by collecting relevance judgements from crowd workers. We randomly chose $300$ enumerating and $300$ counting predicates as returned by our classifiers.
\revision{As co-occurring predicate pairs have a long tail of infrequent pairs that due to small sample size might lead to spurious heuristics scores, we set a threshold on the absolute co-occurrences for the alignments. Of all co-occurring pairs we consider those which co-occur for at least 50 subjects for annotation, evaluation and final ranking purposes.} We then created the set of top-3 counting predicates returned by all the alignment heuristics for each enumerating predicate, so that for each enumerating predicate we had up to \revision{27} counting predicates as candidates. \revision{Note that the alignment is KB-specific, so we return top-3 predicates from the same KB to which the enumerating predicate belongs.}

We repeated the step with the counting predicates, this time returning the top-3 enumerating predicates for each counting predicate. On an average, there were 5 candidates for each set predicate in the enumerating and counting case. The annotation task asked each worker to judge the topical relatedness of a pair of set predicates (an enumerating and a counting predicate) and the degree of completeness based on the integer value of the counting predicate and the entities covered by the enumerating predicate with respect to a subject. 
An example task where the system returns a counting predicate is as follows.

\medskip

\scalebox{0.85}{\hspace{-0.9cm}
\begin{tabular}{llp{3cm}}
     Subject & Predicate & Object\\
     \multicolumn{3}{l}{\textbf{Query}}  \\
     Univ. of California, L.A. & institution$^{-1}$ & Thomas Sowell, Harold Demsetz ..(5 in total)\\
     \multicolumn{3}{l}{\textbf{Result}}\\
     Univ. of California, L.A. & faculty size & 4016
\end{tabular}
}\linebreak

We ask the following two questions.\\
\textit{1. Topical relatedness of \textbf{institution$^{-1}$} to \textbf{faculty size}} is:

\textit{Options: $\circ$ High $\circ$ Moderate $\circ$ Low $\circ$ None}.\linebreak
\textit{2. Enumeration of the objects in the query is:}

\textit{Options: $\circ$ Complete $\circ$ Incomplete $\circ$ Unrelated}.

\medskip

The task in the opposite direction is designed in a similar fashion with the query containing a counting fact and the result, an enumerating fact with the set of objects. 

For this task also we collected three judgements per predicate pair in either direction. We again used a graded relevance system by calculating a mean score of the two responses where, the grades for topical relatedness are \{\textit{High}: 1, \textit{Moderate}: 0.67, \textit{Low}: 0.33, \textit{None}: 0\} and for the completeness of enumeration we have \{\textit{Complete}: 1, \textit{Incomplete}: 0.5, \textit{Unrelated}: 0\}. Thus the graded relevance score (1 being the highest and 0 being the lowest) is calculated by mapping the responses to their grades and averaging over all responses. \revision{Concerning agreement, the pooled standard deviation of responses across pairs was $0.3$ for topical relatedness and $0.46$ for completeness of enumeration.}


\subsection{Classifier models}
We model our classifiers on logistic regression as well as neural networks. However, due to small dataset size, and our interest in interpretable insights, we focus on multiple logistic regression models. We consider a standard \textit{logistic} regression model, a logistic regression model with a weakly informative default \textit{prior}~\cite{gelman2008weakly}, a \textit{Lasso} regularized logistic regression~\cite{tibshirani1996regression} and a \textit{neural} network composed of a hidden layer of size three and sigmoid activation function. Due to the small training set we use Leave-One-Out cross validation to obtain our model performance scores.


All models are compared against a random baseline modelled on the input distribution, i.e., predicting labels at random, with probabilities proportional to label frequency in the training data.

\subsection{Results}

\begin{table}[t]
\centering
\begin{tabular}{|c|ccc|}
\hline
\textbf{Model} & \textbf{Recall} & \textbf{Precision} & \textbf{F1} \\
\hline
Random & 12.8 & 12.8 & 12.8 \\
Logistic & 51.2 & 19.0 & 27.7 \\
Prior & 48.7 & 20.2 & 28.5 \\
Lasso & \textbf{71.7} & \textbf{23.3} & \textbf{35.1} \\
Neural & 35.8 & 20.8 & 26.3 \\
\hline
\end{tabular}
\caption{Performance (precision, recall and F1) of the counting predicate classifier models.}
\label{tbl:prf1_counting}
\end{table}

\begin{table}[t]
\centering
\begin{tabular}{|l|ccc|}
\hline
\textbf{Model} & \textbf{Recall} & \textbf{Precision} & \textbf{F1} \\
\hline
Random & 40.6 & 40.6 & 40.6 \\
Logistic & \textbf{55.6} & 51.7 & 53.5  \\
Prior & \textbf{55.6} & 51.0 & 53.5  \\
Lasso & 51.1 & \textbf{59.6} & \textbf{55.0} \\
Neural & 53.0 & 49.6 & 51.2 \\
\hline
\end{tabular}
\caption{Performance (precision, recall and F1) of the enumerating predicate classifier models.}
\label{tbl:prf1_enumerating}
\end{table}

\paragraph{a. Classifier model selection}
The results of the classifier models are in Tables~\ref{tbl:prf1_counting} and~\ref{tbl:prf1_enumerating}. As one can see, the Lasso regularized model performs the best for counting predicates with an F1 score of $35.1$, which is significantly better than the random model which has an F1 score of $12.8$. We observe that the counting classifier models in general have lower precision scores, but higher recall. The scores of the random model are computed from the training data distribution of counting predicates which contain $39$ positive and $306$ negative datapoints. Note that the number of datapoints is less than the initial selection of $400$ datapoints since, as explained in the previous section, we remove datapoints with divided agreements. 
We use the Lasso regularized model to classify the counting predicates.

In the enumerating predicate scenario also, the Lasso regularized model has an overall highest performance with an F1 score of $55$. Here too, the random classifier performance depends on the distribution of training data which has $133$ positive and $195$ negative datapoints, giving an F1 score of $40.6$. We use the Lasso regularized model for predicting the enumerating predicates. The \revision{comparable recall and precision scores of the enumerating predicate classifier} can be attributed to the almost equal class distribution in the training data, which is not the case for counting predicates.

%
%



\paragraph{b. Important Features}
The most important features in the counting predicate classifier are the mean and 10th percentile of the count values of a predicate with negative weights of $0.006$ and $0.031$ suggesting that counting predicates usually take smaller integer values. The predicate domain of type \textit{Organization} has a positive weight of $0.14$.

The determining features of the enumerating classifier are the type information on the predicate domain and range. For example, the weights for domain \textit{Thing} and range \textit{Organization} are positive values of $0.135$ and $0.046$, respectively. It is interesting to note that predicate ranges of type \textit{Work} and \textit{Place} have small negative weights of $0.008$ and $0.097$, respectively, suggesting that predicates with range type location are less likely to be enumerating predicates.



\begin{table}[t]
    \centering
    \begin{tabular}{|c|c c c|}
        \hline
        \textbf{KB} & \textbf{Input} & \textbf{Output} & \textbf{Filtered}\\
        \hline
         DBP-raw & 16,635  & 4,090 & 4,090 (24.5\%) \\
         DBP-map & 1,670   & 308   & 308 (18.4\%)\\
         WD-truthy & 4,067 & 216   & 203 (4.9\%)\\
         Freebase & 13,872 & 7,752 & 7,614 (54.8\%) \\
         \hline
         \textbf{Total} & 36,244 & 12,366 & 12,215 (33.7\%)\\
         \hline
    \end{tabular}
    \caption{Predicted enumerating predicates across different KBs, where \textbf{Input} is all KB predicates (direct + inverse), \textbf{Output} is from the classifier prediction and \textbf{Filtered} the number of predicates remaining after removing predicates related to IDs and codes.}
    \label{tbl:predicted_ep}
\end{table}

\begin{table}[t]
    \centering
    \begin{tabular}{|c|c c c|}
        \hline
        \textbf{KB} & \textbf{Input} & \textbf{Output} & \textbf{Filtered}\\
         \hline
         DBP-raw & 13,394   & 5,853 & 5,853 (43.6\%) \\
         DBP-map & 1,127    & 898   & 898 (79.6\%)\\
         WD-truthy & 3,346  & 1,922 & 1,067 (31.8\%)\\
         Freebase  & 8,289 & 1,723 & 1,687 (20.3\%)\\
         \hline
         \textbf{Total} & 26,156 & 10,396 & 9,505 (36.3\%)\\
         \hline
    \end{tabular}
    \caption{Predicted counting predicates across different KBs, where \textbf{Input} is KB predicates (direct only), \textbf{Output} is from the classifier prediction and \textbf{Filtered} the number of predicates remaining after removing predicates related to IDs and codes.}
    \label{tbl:predicted_cp}
\end{table}

\begin{table}[t]
\begin{tabular}{|c p{6cm}|}
\hline
\textbf{KB} & \textbf{Counting Predicates} \\ \hline
DBP-raw & \tt{employees}, \tt{retiredNumbers}, \tt{crewMembers}, \tt{postgraduates}, \tt{members} \\
DBP-map & \tt{numberOfStudents}, \tt{facultySize}, \tt{numberOfGoals}, \tt{populationAsOf}, \tt{capacity} \\
WD-truthy & \tt{employees}, \tt{numberOfDeaths}, \tt{numberOfConstituencies}, \tt{numberOfSeats}\\
FB & \tt{children}, \tt{numberOfMembers}, \tt{population}, \tt{numberOfStaff}, \tt{injuries}, \tt{passengers}\\
\hline
\multicolumn{2}{|c|}{\textit{Wrong Predictions}}\\
\hline
DBP-raw & \tt{linecolor}, \tt{km}, \tt{birthyear},\\
DBP-map & \tt{foundingYear}, \tt{keyPerson} \\
WD-truthy & \tt{publicationDate}, \tt{coordinateLocation}\\
FB & \tt{maxLength}, \tt{height}\\
\hline
\end{tabular}
\caption{Example predicted counting predicates from the different KBs. 
}
\label{tbl:predicted_samples_cp}
\end{table}

\paragraph{c. Predicted Set Predicates}
The number of set predicates predicted by each classifier is shown in Tables~\ref{tbl:predicted_ep} and~\ref{tbl:predicted_cp} in the \textit{Output} column. We have $10,396$ predicted as counting predicates by the counting classifier out of $26,156$. The percentage of predicted counting predicates is almost $40\%$ which is much higher than the class distribution in the training data ($11\%$). One reason is the very low precision scores of the classifiers which may lead to more false positives. The enumerating classifier predicts $12,366$ ($34\%$) of $36,244$ predicates as enumerating predicates which is closer to the class distribution seen in the training data ($40\%$). 

\begin{table}[t]
\begin{tabular}{|c p{6.5cm}|}
\hline
\textbf{KB} & \textbf{Enumerating Predicates} \\ \hline
DBP-raw & \tt{college}, \tt{workInstitution}, \tt{affiliations}, \tt{members$^{-1}$}, \tt{voiceActor$^{-1}$}, \tt{nativeLangugae$^{-1}$}, \tt{politicalParty$^{-1}$}\\
DBP-map & \tt{recordLabel}, \tt{developer}, \tt{product}, \tt{publisher}, \tt{formerCoach$^{-1}$} \tt{employer$^{-1}$}, \tt{governor$^{-1}$} \\
WD-truthy & \tt{participantOf$^{-1}$}, \tt{airlineHub}, \tt{developer}, \tt{father$^{-1}$}, \tt{sponsor} \\
FB & \tt{actor}, \tt{member}, \tt{starring}, \tt{publisher}, \tt{airportsServed$^{-1}$}, \tt{foundedLocation$^{-1}$}\\
\hline
\multicolumn{2}{|c|}{\textit{Wrong Predictions}}\\
\hline
DBP-raw & \tt{currentTeam}, \tt{deathCause}, \tt{weightClass} \\
DBP-map & \tt{secondTeam}, \tt{genre}\\
WD-truthy & \tt{parentOrganization}, \tt{hairColor} \\
FB & \tt{cameras}, \tt{burstCapability}, \tt{founder}\\
\hline
\end{tabular},
\caption{Example predicted enumerating predicates from the different KBs.}
\label{tbl:predicted_samples_ep}
\end{table}


We illustrate some predicted counting predicates in Table~\ref{tbl:predicted_samples_cp} and a few enumerating predicates in Table~\ref{tbl:predicted_samples_ep}. The DBpedia raw KB predicate \tt{voiceActor$^{-1}$}, for example, connects a voice actor to the associated shows\footnote{List of shows Mel Blanc voiced over \url{https://tinyurl.com/dbpedia-mel-blanc}} and \tt{employees}\footnote{Sample of subjects with the predicate \tt{employees} \url{https://tinyurl.com/dbpedia-employees}} gives the number of employees in an organization. 

The classifiers also misclassify as shown in previous tables, for example, the counting classifer wrongly predicts dates like \tt{birthYear} and \tt{foundingYear}, measurements such as \tt{km}, \tt{height} as counting predicates. The enumerating classifier makes errors by positively labelling functional and pseudo-functional predicates like \tt{currentTeam}, \tt{sourceOfIncome}.

\begin{table}[t]
\centering
\begin{tabular}{|c | c c | c|}
\hline
\textbf{Class} & \textbf{Pre-filter} & \textbf{Post-filter} & \textbf{Classifier} \\ \hline
Enumerating & 2,167 & 151 & 2,016 ($93\%$) \\
Counting & 2,158 & \revision{881} & 1,277 ($59\%$) \\
\hline
\end{tabular}
\caption{The number of identifier predicates present in the input to the classifiers \textbf{Pre-filter} vs.\ the number present in the predicted predicates \textbf{Post-filter} successfully removed by the classifiers.}
\label{tbl:label_filter}
\end{table} 

\paragraph{d. Filtering identifier labels}
Our classifiers, especially the counting classifier has lower precision than recall. One of the commonly occurring type of predicates are identifiers, which may be represented as a fact with an large number in integer or string format and, we can remove such predicates without losing any actual set predicate.  The filtering is done by checking for the presence of the words \textit{`id'} and \textit{`code'} as substrings, but not part of a longer word, in the predicate label, irrespective of the source KB.
In Table~\ref{tbl:label_filter} we compare the number of identifier predicates that need to be filtered before classification versus the number of predicates filtered after classification. The enumerating classifier is good at filtering identifier predicates since almost $90\%$ of the identifier labels are predicted to be false. The counting classifier removes around $59\%$ of the identifier predicates and could benefit from the identifier filter. Thus we apply the identifier label filter only on the output of the classifiers, and the final numbers are shown in the \textit{Filtered} column of Tables~\ref{tbl:predicted_ep} and~\ref{tbl:predicted_cp}.


\begin{table}[t]
    \centering
    \scalebox{0.9}{
    \begin{tabular}{|l|C{1cm} C{1cm} | C{1cm} C{1cm}|}
    \hline
    \multirow{2}{*}{\textbf{Metric}} & \multicolumn{2}{c|}{\textbf{Counting}} & \multicolumn{2}{c|}{\textbf{Enumerating}}\\
     & $@1$ & $@3$ & $@1$ & $@3$ \tabularnewline
     \hline
     \textit{Absolute} & 0.71 & 0.56 & 0.62 & 0.63 \tabularnewline
     \textit{Jaccard} & 0.76 & 0.61 & 0.69 & 0.67 \tabularnewline
     \textit{Conditional$_C$} & 0.71 & 0.56 & 0.68 & 0.67 \tabularnewline
     \textit{Conditional$_E$} & 0.76 & 0.68 & 0.62 & 0.63 \tabularnewline
     \textit{P'wiseMI} & 0.73 & 0.58 & 0.71 & 0.70 \tabularnewline 
     \hline
     \textit{P'fectMR} & 0.70 & 0.57 & 0.73 & 0.72 \tabularnewline
     \textit{Correlation} & 0.77 & 0.69 & 0.62 & 0.61 \tabularnewline
     \textit{P'tileVM} & 0.72 & 0.57 & 0.65 & 0.65  \tabularnewline
     \hline
     \textit{CosineSim} & 0.79 & 0.61 & 0.74 & 0.73 \tabularnewline
     \hline
     \textbf{\textit{Combined}} & \textbf{0.84} & \textbf{0.67} & \textbf{0.75} & \textbf{0.75} \tabularnewline
     \hline
    \end{tabular}
    } 
    \caption{Average \textit{NDCG} scores for the alignment stage.}
    \label{tbl:ndcg}
\end{table}

\paragraph{e. Statistical alignment} 
%
%
The \textit{NDCG} scores reported in Table~\ref{tbl:ndcg} are an evaluation of the top three alignments from all nine alignment metrics based on relevance judgments collected from crowd workers. We report the \textit{NDCG} at positions 1 and 3. The table is divided into the three alignment families and we consider two directions. The first is the direction from a counting predicate to its enumerating predicate alignments and the second is the reverse.

Based on the scores presented in Table~\ref{tbl:ndcg}, we can conclude that the linguistic similarity metric of cosine similarity \revision{(defined in Sec.~\ref{par:metric_ling}C)} performs the best individually, except for \textit{NDCG@3} for the counting to enumerating direction, where the Pearson correlation measure performs best. 
The \textit{Correlation} metric in the counting to enumerating direction and the \textit{P'fectMR} metric in the reverse direction are the best performing metrics of the set predicate value distribution family \revision{(defined in Sec.~\ref{par:metric_vald}B)}. The strongest metrics in the set predicate co-occurrence family \revision{(defined in Sec.~\ref{par:metric_cooccur}A)} are \textit{Conditional$_E$} in the direction of counting to enumerating predicate alignment and \textit{P'wiseMI} in the other direction.

The \textit{Combined} metric takes the best performing metric from each family and computes the mean of the alignment scores to obtain a combined score which gives better results than any individual metric. We use this combined measure to rank our alignments. 



\begin{table}[t]
\centering
\begin{tabular}{|l|cccc|}
\hline
& {DBP-raw} & {DBP-map} & {WD-truthy} & {FB} \\
\hline
DBP-raw & 57.1 & 36.0 & 17.6 & - \\
DBP-map & 58.5 & 52.5 & 50.0 & -\\
WD-truthy & 54.7 & 51.4 & 52.0 & -\\
FB & 41.4 & 41.8 & 8.5 & 80.0\\
Random & 39.4 & 33.3 & 33.3 & 60.8 \\
\hline
\end{tabular}
\caption{F1 scores for enumerating predicate classifiers, where rows heads represent the training sets and the column heads represent the test sets.}
\label{tbl:transfer_enum}
\end{table}

\begin{table}[t]
\centering
\begin{tabular}{|l|cccc|}
\hline
& {DBP-raw} & {DBP-map} & {WD-truthy} & {FB} \\
\hline
DBP-raw & 56.5 & 5.2 & 33.3 & - \\
DBP-map & 15.1 & 43.7 & 11.7 & 12.1\\
WD-truthy & 25.9 & 20.4 & 32.0 & 8.5\\
FB & 34.6 & 13.7 & 7.4 & 40.0\\
Random & 18.7 & 11.1 & 14.2 & 14.2 \\
\hline
\end{tabular}
\caption{F1 scores for counting predicate classifiers, where rows heads represent the training sets and the column heads represent the test sets.}
\label{tbl:transfer_count}
\end{table}

\paragraph{f. Transferability}
\revision{We also evaluate the transferability of set predicate identification models trained on one KB, evaluated on the others. We do this for all combinations of KBs and report the F1 scores for the enumerating predicates in Table~\ref{tbl:transfer_enum} and the counting predicates in Table~\ref{tbl:transfer_count}. In each setting we also report the F1 scores of a random baseline.}

\revision{We observe that in most cases, the classifiers significantly outperform the random baseline, although the performance is quite low when evaluating on the Wikidata-truthy and Freebase KBs. The task of predicting enumerating predicates in the two variants of DBpedia is better when trained on any of the remaining KBs, whereas no classifier can predict enumerating predicates in Freebase. In the counting predicate prediction, the classifier trained on DBpedia-map performs worse than random classifier for all test data. Here too, the performance of classifiers on Freebase is quite low. In contrast, the classifiers trained on Freebase perform well on DBpedia in both cases. We can conclude that the case of counting predicate prediction is more challenging, given that the F1 scores of the random classifier and the training data are lower. Additionally, training on a single KB does not fare well in the case of counting predicate prediction.}

\section{Use Cases}

\paragraph{Question answering}
\begin{figure}[t]
 \centering 
 \includegraphics[width=0.46\textwidth]{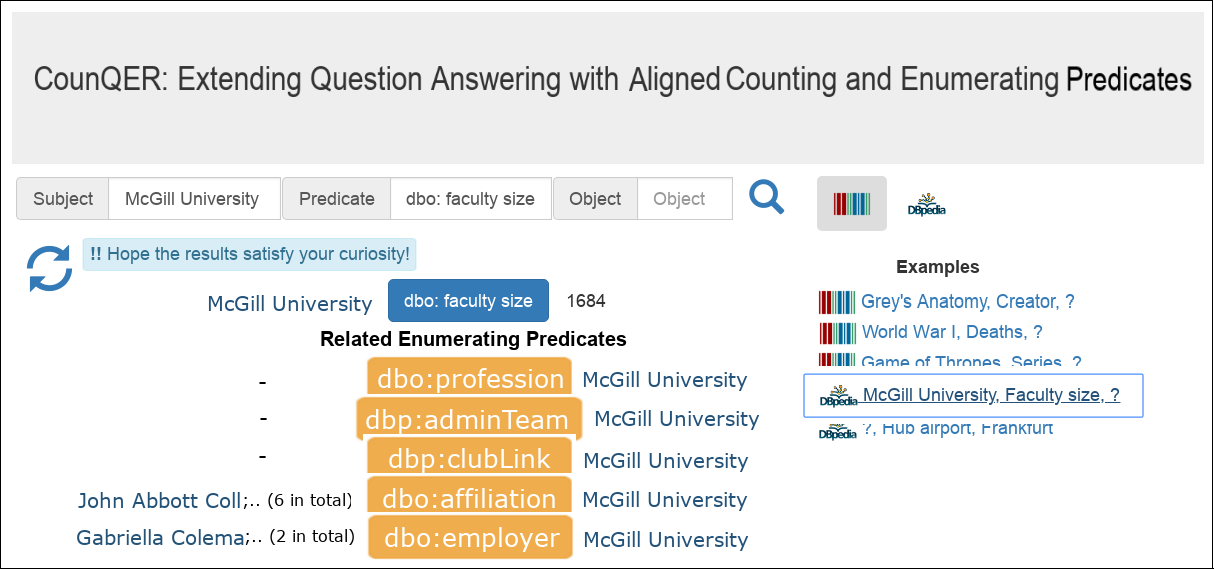}
\caption{Demo system for question answering. 
}
 \label{fig:demo}
\end{figure}
\revision{Question answering benefits from set predicate alignment at three stages: (i) natural language parsing, (ii) query result debugging, (iii) query result enrichment.}

\revision{In \emph{natural language parsing}, state-of-the-art KBQA systems typically generate a set of candidate parses, which subsequently are ranked, and the top one executed (with research prototypes often allowing the user to inspect and choose among them). Set predicate alignments could be used to generate further candidates, or used as ranking feature.}

\revision{Structured queries over KBs often produce empty or otherwise unexpected results that require \emph{query debugging}. Set predicate alignments could help to understand whether an ambiguous predicate name refers to the intended predicate, and related counts could help to understand that a predicate is an intended one, but that just the KB is incomplete.}

\revision{Finally, in the absence of any problems, results from aligned predicates can \emph{enhance the user experience} and anticipate follow-up information needs.  Given a query such as \emph{``SELECT ?val WHERE \{dbp:McGill\_University dbo:faculty\_size ?val.\}''}, we can, in addition to the actual count, directly provide example instance (see Fig.~\ref{fig:demo}).\footnote{Google Web Search shows this behaviour for queries such as ``number of movies by Tarantino''.} Conversely, if the user asks for people working at McGill University, we can, in addition to instances, provide (an estimate of) the total count.}

We are in the process of building a demo for that purpose. In the demo interface, users can input a single-triple Wikidata or DBpedia query by specifying a predicate and a subject or an object. Along with the results for the other field, the interface will then show a ranked list of aligned set predicates, both, those having values (bottom 2 in Fig.~\ref{fig:demo}), and those having no values (top 3 in same Figure). This makes the demo relevant for two use cases: (i) KB curation, by checking which related predicates have missing information so far, and (ii) question answering, by enhancing count questions with instance information, and vice versa. 
A video of the demo prototype can be found at \url{https://tinyurl.com/y2ka4kfu}, and the demo can be accessed at  \url{https://counqer.mpi-inf.mpg.de/spo}.

\begin{figure}[t]
 \centering 
 \includegraphics[width=0.46\textwidth]{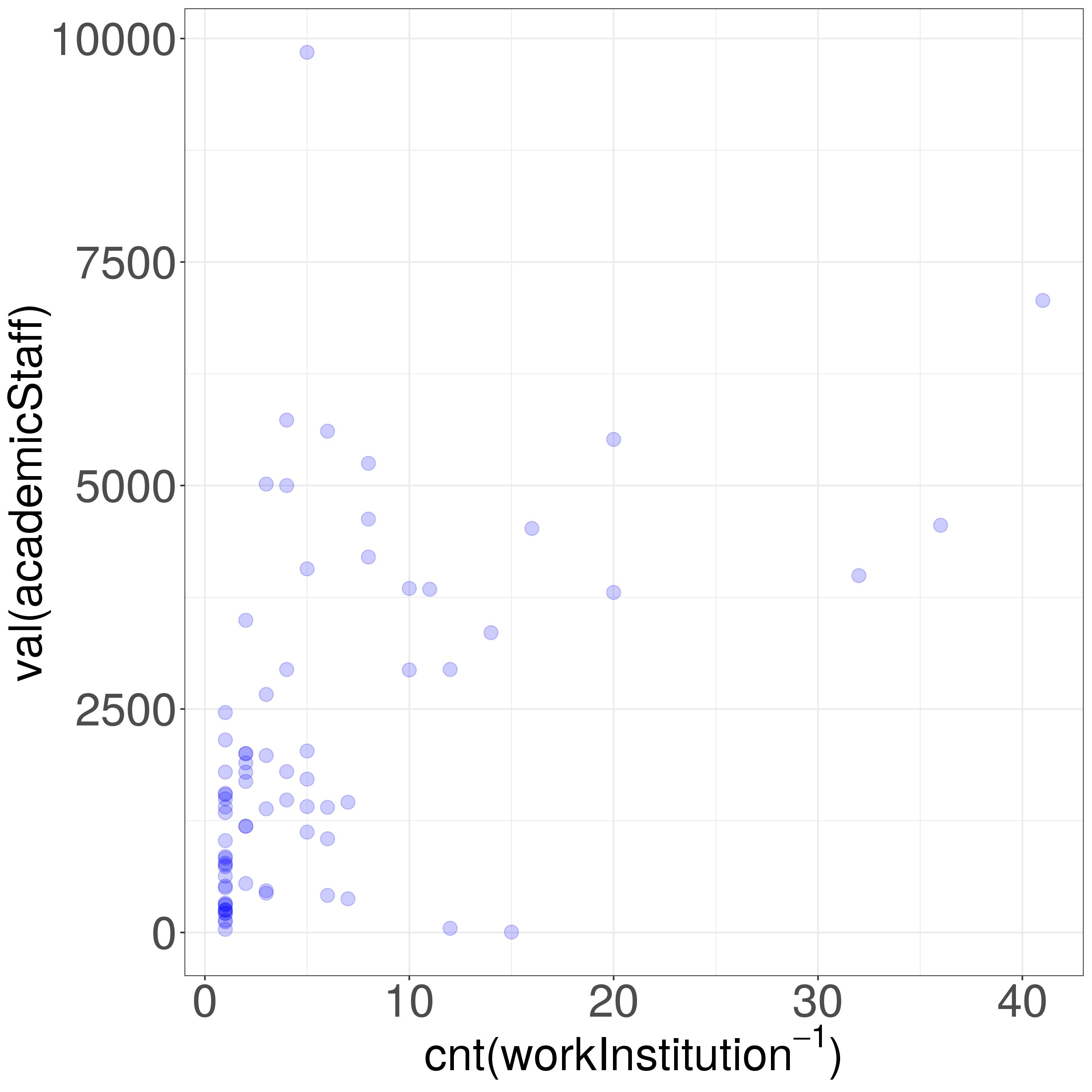}
\caption{Value distribution of \tt{academicStaff} and count of \tt{workInstitution$^{-1}$} across 76 subjects in DBpedia raw KB.}
 \label{fig:align_dbp_raw1}
\end{figure}

\paragraph{KB curation}
In this section we look into a few alignments from different KBs and the distribution of their values. \revision{The first alignment in Fig.~\ref{fig:align_dbp_raw1} is the pair (\tt{work\linebreak~Institution$^{-1}$}, \tt{academicStaff}) from the DBpedia raw KB, which co-occurs across 76 subjects (institutions). 
Each point $(x,y)$ in the plot represents an institution, which is connected to $x$ entities by the predicate \tt{workInstitution$^{-1}$}, and which takes the value y for the predicate \tt{academicStaff}. In an ideal condition the count of instances should match the value and all points should lie along the line $y=x$. Points lying above this line suggest incompleteness. Such is the case in Fig.~\ref{fig:align_dbp_raw1} where the predicate \tt{workInstitution$^{-1}$} is often only connected to the popular or important staff members.}

\begin{figure}[t]
 \centering 
 \includegraphics[width=0.46\textwidth]{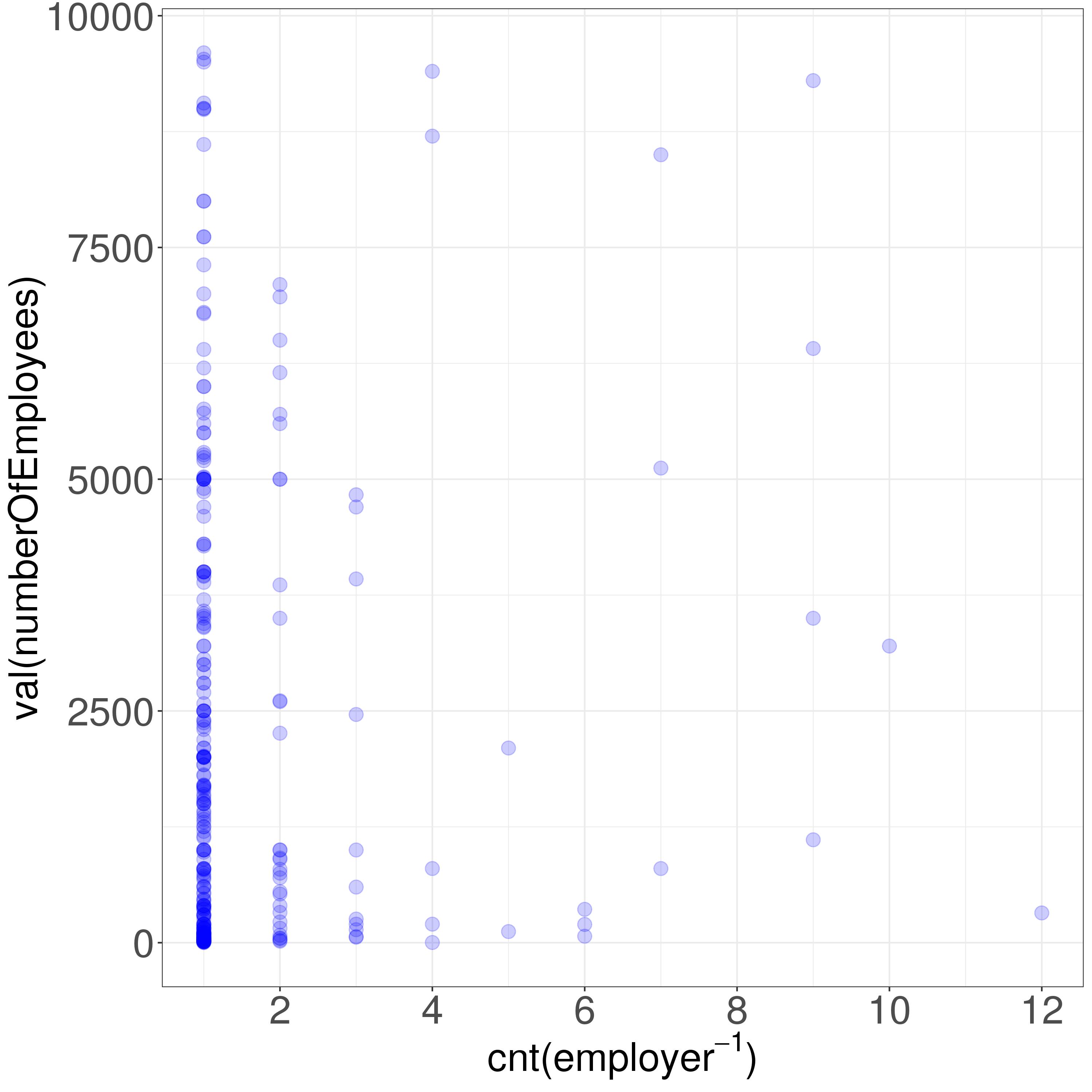}
\caption{Value distribution of counting predicate \tt{numberOfEmployees} and count of enumerating predicate \tt{employer$^{-1}$} across 278 subjects in DBpedia mapping-based KB.}
 \label{fig:align_dbp_map1}.
\end{figure}

Next we look into an alignment from DBpedia mapping based KB, (\tt{employer$^{-1}$}, \tt{numberOfEmployees}) in Fig.~\ref{fig:align_dbp_map1}. In this alignment we also observe that the enumerated facts is much smaller than the number of employees, typically because such facts only exist for the most important employees.

\begin{figure}[t]
 \centering 
 \includegraphics[width=0.46\textwidth]{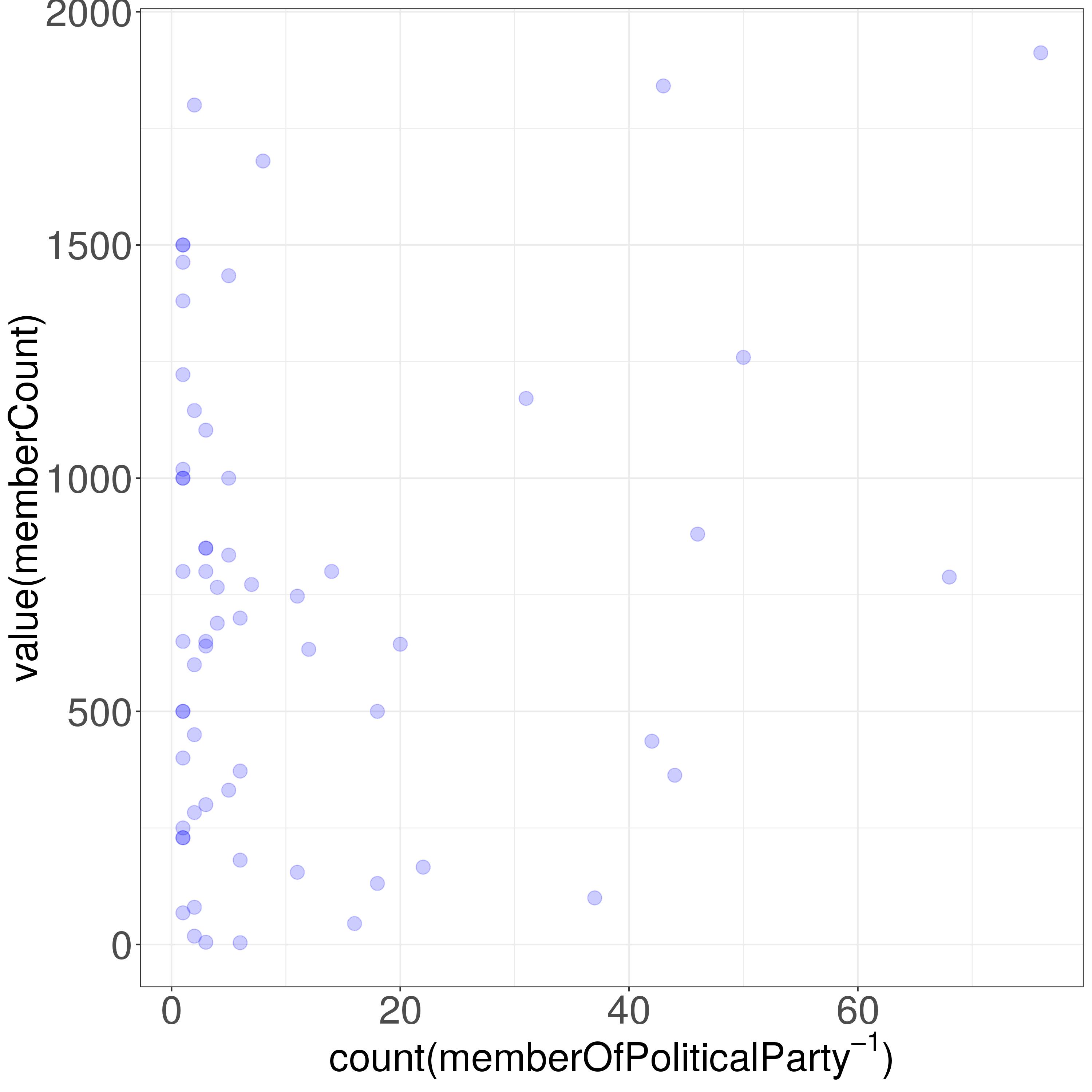}
\caption{Value distribution of counting predicate \tt{memberCount} and count of enumerating predicate \tt{memberOfPoliticalParty$^{-1}$} across 62 subjects in Wikidata-truthy KB.}
 \label{fig:align_wd1}.
\end{figure}

\begin{figure}[t]
 \centering 
 \includegraphics[width=0.47\textwidth]{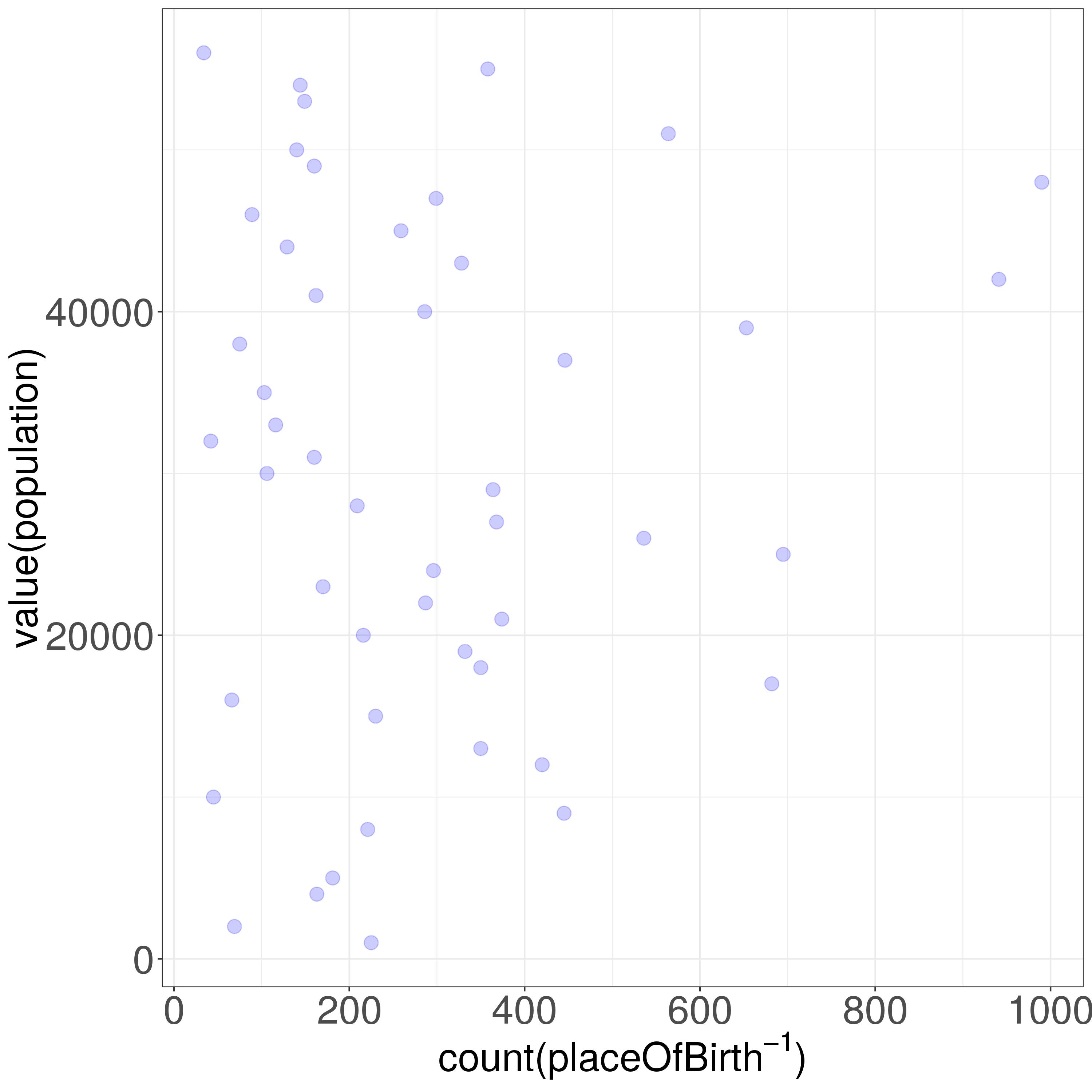}
\caption{Value distribution of counting predicate \tt{populationState} and count of enumerating predicate \tt{placeOfBirth$^{-1}$} across 48 subjects in Freebase KB.}
 \label{fig:align_fb2}.
\end{figure}

In Fig.~\ref{fig:align_wd1}, we show an alignment from the Wikidata KB which is regarding the members of a political party (\tt{memberOfPoliticalParty$^{-1}$}, \tt{memberCount}). Similar to the previous trends, here also the number of enumerated facts about the members in a political party is less than the actual value. The final alignment we show is of the pair (\tt{placeOfBirth$^{-1}$, \tt{populationState}}) in the Freebase KB as shown in Fig.~\ref{fig:align_fb2}. From the numbers it seems that the predicate covers small geographical locations where the number of enumerated facts is more complete than in the previous cases. 

\revision{In each of the alignment figures there are at most two instances where the value of the counting predicate is less than the count of enumerated instances, there exist no such instances in the Fig.~\ref{fig:align_fb2}. Such a low number of points indicate anomaly or inconsistencies such as a backdated counting predicate value or, for instance, in Fig.~\ref{fig:align_dbp_map1} where the value of counting predicate \tt{numberofEmployees} is $0$ even though there exist enumerated facts.}


\begin{figure*}[t]
 \centering 
 \includegraphics[width=0.8\textwidth]{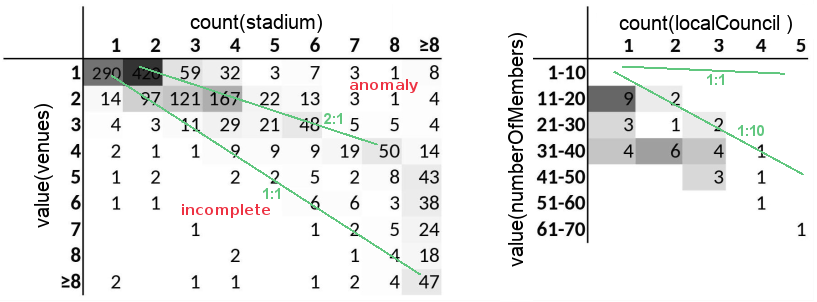}
\caption{Value distribution of the counting predicate \tt{venues} and count of enumerating predicate \tt{stadium} for 2179 sports events (left), and \tt{numberOfMembers} and count of \tt{localCouncil}$^{-1}$ for 35 political assemblies (right) from the DBpedia raw KB.}
 \label{tbl:heatmaps}
\end{figure*}



Figure~\ref{tbl:heatmaps} shows the value distribution of an alignment where each cell in position ($x,y$) gives the number of subjects which takes $x$ as the value of the counting predicate and $y$ number of instances with the enumerating predicate. The first analysis concerns the places where a sports event took place. A notable anomaly in DBpedia raw KB is that regularly, for each venue, both the stadium and the city were recorded. Thus, we plot two green lines showing 1:1 matches, and 2:1 matches. Instances below both lines likely point to incompleteness (some stadiums are missing), instances below both lines likely point to some errors in the data (i.e., too many stadiums added). As one can see, the completeness appears to be relatively high, while there are several cases that deserve closer inspection w.r.t.\ possible incorrectness.

The second analysis concerns the number of members of local councils, compared with individual members listed. Here incompleteness is prevalent, with typically only 1 in 30 to 1 in 10 members listed.

\section{Discussion}

\paragraph{Limitations and usage implications}
\revisionsimon{While the experimental results are encouraging, F1-scores of $35-55\%$ for the identification stage indicate that there is still much room for improvement. Further feature engineering, e.g., regarding IDs and codes, and further distribution measures, may still yield moderate improvements, but we see three principled challenges that limit any fully-automated approach: (i) statistical cues allow to rule out well some clear negatives, but for many infrequent predicates, provide only weak signals. (ii) textual cues for predicates are mostly short and thus of limited informativeness. (iii) input KBs come with a considerable level of noise.}

\revisionsimon{Where possible, we would therefore recommend to execute set predicate identification and alignment not in a fully automated manner, but employ a human-in-the-loop process, where statistical procedures narrow the search space for human annotators, or human annotators focus on the ``fat head'' (e.g., the 100 most used predicates), and automated methods focus on the long tail. In industrial deployment of the question answering use case, human input could also come from user feedback, e.g., query-click-logs or query reformulations and follow-up questions.}



\paragraph{Inverse predicates}
\revision{For generalizability, our method currently does not incorporate existing definitions of inverse predicates.}  
\revision{Freebase contains $12K$ such definitions (via \texttt{owl:inverseOf}~\cite{chah2018ok}), Wikidata  has $136$ (identified via the meta-property \emph{inverse property}), DBpedia has three mentions in comment fields. Incorporating such KB-specific constraints could further boost the accuracy of alignments.} 

\paragraph{Transferability of method}
A crucial aspect for our framework is whether it can be utilized on new KBs without requiring too much adaptation. Our modular framework is aimed towards this purpose.
The supervised predicate classification stage allows to transfer our approach by only creating new training instances. For KBs where textual predicates are unavailable, a sensible extension is the incorporation of latent representations of predicates~\cite{wang2014knowledge,lin2015learning} \revision{as separate features in the classification stage and by considering cosine similarity of predicate embeddings as a heuristic in the predicate alignment.}

\paragraph{Indirect alignments} 
The alignments are also helpful in identifying redundancies in schema, where two or more set predicates (enumerating/counting) describing the same concept exist and are all aligned to a single set predicate of the other variant. For example the enumerating predicate \tt{affiliation} in the DBpedia mapping-based KB aligns with the counting predicates \{\tt{facultySize}, \tt{staff}, \tt{numberOfStaff}\}. 

\paragraph{Multi-hop alignments} 
Counting predicates may well align with multi-hop paths of enumerating predicates. For instance, an interesting near-subset of \tt{population(x,y)} is \tt{worksAt(y,z),\! basedIn(z,x)}. The search space for such alignments would grow quadratically, but clever pruning may keep it manageable.



\paragraph{Crowd annotation costs}
The cost of the annotating classifier training data is approximately $0.13\$$ per task per judgment. For the alignment evaluation task the cost was almost $0.40\$$ per task per judgment. Thus the average cost per task is $0.5(0.13*3 + 0.4*3) \approx 0.80\$$ if we collect 3 judgments. \revision{Given the average time spent per task, we arrive at an hourly pay of \$14, which corresponds to the salary of student assistants in Germany.} 

\paragraph{Open information extraction}
So far we have only considered the alignment of canonicalized KB predicates. An interesting direction would be to extend this alignment towards open information extraction and open knowledge bases in the style of Reverb~\cite{reverb}, i.e., to align textual phrases like \emph{``X has Y employees''} with phrases like \emph{``Z works at Y'', ``Z recently joined X''}, etc. Numeric open information extraction traditionally focuses on temporal information~\cite{ling2010temporal} and measures~\cite{saha2017bootstrapping}, though there are also some recent works on counting information extraction~\cite{mirza2017cardinal,mirza2018enriching}, which one might build upon.

\section{Conclusion}

In this paper we have introduced the problem of set predicate alignment, and presented CounQER, a methodology for identifying and linking set predicates that combines co-occurrence, correlational and linguistic features.
\revisionsimon{We have shown that automated methods can identify and align set predicates on four diverse knowledge bases, and that these alignments are useful for use cases in knowledge base curation and question answering.}

\revisionsimon{We believe that understanding set predicate semantics in today's KBs is an important step towards a better interaction with structured world knowledge repositories.}
Our next goals are to extend this methodology to multi-hop alignments, and towards open predicate phrases extracted from natural language texts.

\bibliographystyle{splncs04}
\bibliography{refs}




\end{document}